\documentclass[aps,prd,preprint,superscriptaddress,showpacs]{revtex4-1}

\usepackage{graphicx}

\usepackage{grffile}
\usepackage{bm} \usepackage[update,prepend]{epstopdf}
\usepackage[usenames,dvipsnames,svgnames,table]{xcolor}
\usepackage{lineno}
\usepackage{xspace}
\usepackage{subfigure}
\usepackage{caption}
\usepackage{hyperref}
\usepackage[noabbrev,capitalize]{cleveref}
\usepackage{hyphenat}

\usepackage{wrapfig}
\usepackage{changebar}
\cbcolor{yellow}

\usepackage{dcolumn}

\newcommand{\myref}[5]{ #1, #3 {\bf #4}, #5 (#2). }
\newcommand{\arxivref}[5]{ #1, #5 (#2). }

\mathchardef\mhyphen="2D

\newcommand{\fiducialmass}{\ensuremath{0.9\textrm{~t}}}

\newcommand{\bestFitMaccqeResultShapeOnly}{\ensuremath{1.43^{+0.28}_{-0.22} \textrm{ GeV}/c^{2}}}
\newcommand{\bestFitMaccqeResultShapePlusNorm}{\ensuremath{1.26^{+0.21}_{-0.18} \textrm{ GeV}/c^{2}}}

\newcommand{\numu}{\ensuremath{\nu_{\mu}}}
\newcommand{\nue}{\ensuremath{\nu_{e}}}
\newcommand{\numubar}{\ensuremath{\bar{\nu}_{\mu}}}
\newcommand{\nuebar}{\ensuremath{\bar{\nu}_{e}}}

\newcommand{\eNu}{\ensuremath{E_{\nu}}}
\newcommand{\eNuTruth}{\ensuremath{E_{\nu}^{\mathrm{true}}}}

\newcommand{\pMu}{\ensuremath{p_{\mu}}}
\newcommand{\thetaMu}{\ensuremath{\theta_{\mu}}}
\newcommand{\cosThetaMu}{\ensuremath{\mathrm{cos}\theta_{\mu}}}
\newcommand{\pMuCosThetaMu}{\ensuremath{\pMu\textrm{-}\cosThetaMu}}
\newcommand{\xY}{\ensuremath{x\mhyphen y}}

\newcommand{\maccqe}{\ensuremath{M_{A}^{\mathrm{QE}}}}

\newcommand{\maccqeNorm}{\ensuremath{\maccqe\textrm{-}\mathrm{norm}}}
\newcommand{\maccqeShape}{\ensuremath{\maccqe\textrm{-}\mathrm{shape}}}

\newcommand{\mares}{\ensuremath{M_{A}^{\mathrm{RES}}}}
\newcommand{\pF}{\ensuremath{p_{F}}}

\newcommand{\pizero}{\ensuremath{\pi^{0}}}

\newcommand{\GeV}{~\ensuremath{\mathrm{GeV}}\xspace}

\newcommand{\GeVoverCSq}{~\ensuremath{\mathrm{GeV}/c^{2}}\xspace}
\newcommand{\GeVoverC}{~\ensuremath{\mathrm{GeV}/c}\xspace}

\newcommand{\mm}{~\mathrm{mm}\xspace}

\newcommand{\nanosec}{~\mathrm{ns}\xspace}

\newcommand{\avgEnu}{\ensuremath{\langle \eNu \rangle}}
\newcommand{\fluxAvgSigma}{\langle \sigma \rangle} 
\newcommand{\NumPredictedBinI}{\ensuremath{N_{i}^{\mathrm{predicted}}(w_{jk}, d_{i}, f_{j}, x_{ijk}) }} 
 
\newcommand{\NumPredictedBinITheta}{\ensuremath{N_{i}^{\mathrm{predicted}}(\bm{\Theta}) }} 
\newcommand{\NumObservedBinITheta}{\ensuremath{N_{i}^{\mathrm{observed}}}}

\newcommand{\CCQE}{\ensuremath{\mathrm{CCQE}}}

\newcommand{\cmSq}{~\ensuremath{\mathrm{cm}^{2}}\xspace}
\newcommand{\fluxAvgXsecNominalNEUT}{\ensuremath{0.88 \times 10^{-38}}\cmSq}

\newcommand{\fluxAvgXsecRFG}{\ensuremath{(0.83 \pm 0.12) \times 10^{-38}}\cmSq}

\newcommand{\collcom}[1]{#1}
\newcommand{\revcom}[1]{#1}
 
\begin{document}

        \title[Measurement of the $\nu_\mu$ CCQE cross section with ND280 at T2K]{Measurement of the $\nu_\mu$ CCQE cross section on carbon with the ND280 detector at T2K}
    \date{\today}
    \newcommand{\INSTC}{\affiliation{University of Alberta, Centre for Particle Physics, Department of Physics, Edmonton, Alberta, Canada}}
\newcommand{\INSTEE}{\affiliation{University of Bern, Albert Einstein Center for Fundamental Physics, Laboratory for High Energy Physics (LHEP), Bern, Switzerland}}
\newcommand{\INSTFE}{\affiliation{Boston University, Department of Physics, Boston, Massachusetts, U.S.A.}}
\newcommand{\INSTD}{\affiliation{University of British Columbia, Department of Physics and Astronomy, Vancouver, British Columbia, Canada}}
\newcommand{\INSTGA}{\affiliation{University of California, Irvine, Department of Physics and Astronomy, Irvine, California, U.S.A.}}
\newcommand{\INSTI}{\affiliation{IRFU, CEA Saclay, Gif-sur-Yvette, France}}
\newcommand{\INSTGB}{\affiliation{University of Colorado at Boulder, Department of Physics, Boulder, Colorado, U.S.A.}}
\newcommand{\INSTFG}{\affiliation{Colorado State University, Department of Physics, Fort Collins, Colorado, U.S.A.}}
\newcommand{\INSTFH}{\affiliation{Duke University, Department of Physics, Durham, North Carolina, U.S.A.}}
\newcommand{\INSTBA}{\affiliation{Ecole Polytechnique, IN2P3-CNRS, Laboratoire Leprince-Ringuet, Palaiseau, France }}
\newcommand{\INSTEF}{\affiliation{ETH Zurich, Institute for Particle Physics, Zurich, Switzerland}}
\newcommand{\INSTEG}{\affiliation{University of Geneva, Section de Physique, DPNC, Geneva, Switzerland}}
\newcommand{\INSTDG}{\affiliation{H. Niewodniczanski Institute of Nuclear Physics PAN, Cracow, Poland}}
\newcommand{\INSTCB}{\affiliation{High Energy Accelerator Research Organization (KEK), Tsukuba, Ibaraki, Japan}}
\newcommand{\INSTED}{\affiliation{Institut de Fisica d'Altes Energies (IFAE), Bellaterra (Barcelona), Spain}}
\newcommand{\INSTEC}{\affiliation{IFIC (CSIC \& University of Valencia), Valencia, Spain}}
\newcommand{\INSTEI}{\affiliation{Imperial College London, Department of Physics, London, United Kingdom}}
\newcommand{\INSTGF}{\affiliation{INFN Sezione di Bari and Universit\`a e Politecnico di Bari, Dipartimento Interuniversitario di Fisica, Bari, Italy}}
\newcommand{\INSTBE}{\affiliation{INFN Sezione di Napoli and Universit\`a di Napoli, Dipartimento di Fisica, Napoli, Italy}}
\newcommand{\INSTBF}{\affiliation{INFN Sezione di Padova and Universit\`a di Padova, Dipartimento di Fisica, Padova, Italy}}
\newcommand{\INSTBD}{\affiliation{INFN Sezione di Roma and Universit\`a di Roma ``La Sapienza'', Roma, Italy}}
\newcommand{\INSTEB}{\affiliation{Institute for Nuclear Research of the Russian Academy of Sciences, Moscow, Russia}}
\newcommand{\INSTHA}{\affiliation{Kavli Institute for the Physics and Mathematics of the Universe (WPI), The University of Tokyo Institutes for Advanced Study, University of Tokyo, Kashiwa, Chiba, Japan}}
\newcommand{\INSTCC}{\affiliation{Kobe University, Kobe, Japan}}
\newcommand{\INSTCD}{\affiliation{Kyoto University, Department of Physics, Kyoto, Japan}}
\newcommand{\INSTEJ}{\affiliation{Lancaster University, Physics Department, Lancaster, United Kingdom}}
\newcommand{\INSTFC}{\affiliation{University of Liverpool, Department of Physics, Liverpool, United Kingdom}}
\newcommand{\INSTFI}{\affiliation{Louisiana State University, Department of Physics and Astronomy, Baton Rouge, Louisiana, U.S.A.}}
\newcommand{\INSTJ}{\affiliation{Universit\'e de Lyon, Universit\'e Claude Bernard Lyon 1, IPN Lyon (IN2P3), Villeurbanne, France}}
\newcommand{\INSTHB}{\affiliation{Michigan State University, Department of Physics and Astronomy,  East Lansing, Michigan, U.S.A.}}
\newcommand{\INSTCE}{\affiliation{Miyagi University of Education, Department of Physics, Sendai, Japan}}
\newcommand{\INSTDF}{\affiliation{National Centre for Nuclear Research, Warsaw, Poland}}
\newcommand{\INSTFJ}{\affiliation{State University of New York at Stony Brook, Department of Physics and Astronomy, Stony Brook, New York, U.S.A.}}
\newcommand{\INSTGJ}{\affiliation{Okayama University, Department of Physics, Okayama, Japan}}
\newcommand{\INSTCF}{\affiliation{Osaka City University, Department of Physics, Osaka, Japan}}
\newcommand{\INSTGG}{\affiliation{Oxford University, Department of Physics, Oxford, United Kingdom}}
\newcommand{\INSTBB}{\affiliation{UPMC, Universit\'e Paris Diderot, CNRS/IN2P3, Laboratoire de Physique Nucl\'eaire et de Hautes Energies (LPNHE), Paris, France}}
\newcommand{\INSTGC}{\affiliation{University of Pittsburgh, Department of Physics and Astronomy, Pittsburgh, Pennsylvania, U.S.A.}}
\newcommand{\INSTFA}{\affiliation{Queen Mary University of London, School of Physics and Astronomy, London, United Kingdom}}
\newcommand{\INSTE}{\affiliation{University of Regina, Department of Physics, Regina, Saskatchewan, Canada}}
\newcommand{\INSTGD}{\affiliation{University of Rochester, Department of Physics and Astronomy, Rochester, New York, U.S.A.}}
\newcommand{\INSTBC}{\affiliation{RWTH Aachen University, III. Physikalisches Institut, Aachen, Germany}}
\newcommand{\INSTFB}{\affiliation{University of Sheffield, Department of Physics and Astronomy, Sheffield, United Kingdom}}
\newcommand{\INSTDI}{\affiliation{University of Silesia, Institute of Physics, Katowice, Poland}}
\newcommand{\INSTEH}{\affiliation{STFC, Rutherford Appleton Laboratory, Harwell Oxford,  and  Daresbury Laboratory, Warrington, United Kingdom}}
\newcommand{\INSTCH}{\affiliation{University of Tokyo, Department of Physics, Tokyo, Japan}}
\newcommand{\INSTBJ}{\affiliation{University of Tokyo, Institute for Cosmic Ray Research, Kamioka Observatory, Kamioka, Japan}}
\newcommand{\INSTCG}{\affiliation{University of Tokyo, Institute for Cosmic Ray Research, Research Center for Cosmic Neutrinos, Kashiwa, Japan}}
\newcommand{\INSTGI}{\affiliation{Tokyo Metropolitan University, Department of Physics, Tokyo, Japan}}
\newcommand{\INSTF}{\affiliation{University of Toronto, Department of Physics, Toronto, Ontario, Canada}}
\newcommand{\INSTB}{\affiliation{TRIUMF, Vancouver, British Columbia, Canada}}
\newcommand{\INSTG}{\affiliation{University of Victoria, Department of Physics and Astronomy, Victoria, British Columbia, Canada}}
\newcommand{\INSTDJ}{\affiliation{University of Warsaw, Faculty of Physics, Warsaw, Poland}}
\newcommand{\INSTDH}{\affiliation{Warsaw University of Technology, Institute of Radioelectronics, Warsaw, Poland}}
\newcommand{\INSTFD}{\affiliation{University of Warwick, Department of Physics, Coventry, United Kingdom}}
\newcommand{\INSTGE}{\affiliation{University of Washington, Department of Physics, Seattle, Washington, U.S.A.}}
\newcommand{\INSTGH}{\affiliation{University of Winnipeg, Department of Physics, Winnipeg, Manitoba, Canada}}
\newcommand{\INSTEA}{\affiliation{Wroclaw University, Faculty of Physics and Astronomy, Wroclaw, Poland}}
\newcommand{\INSTH}{\affiliation{York University, Department of Physics and Astronomy, Toronto, Ontario, Canada}}

\INSTC
\INSTEE
\INSTFE
\INSTD
\INSTGA
\INSTI
\INSTGB
\INSTFG
\INSTFH
\INSTBA
\INSTEF
\INSTEG
\INSTDG
\INSTCB
\INSTED
\INSTEC
\INSTEI
\INSTGF
\INSTBE
\INSTBF
\INSTBD
\INSTEB
\INSTHA
\INSTCC
\INSTCD
\INSTEJ
\INSTFC
\INSTFI
\INSTJ
\INSTHB
\INSTCE
\INSTDF
\INSTFJ
\INSTGJ
\INSTCF
\INSTGG
\INSTBB
\INSTGC
\INSTFA
\INSTE
\INSTGD
\INSTBC
\INSTFB
\INSTDI
\INSTEH
\INSTCH
\INSTBJ
\INSTCG
\INSTGI
\INSTF
\INSTB
\INSTG
\INSTDJ
\INSTDH
\INSTFD
\INSTGE
\INSTGH
\INSTEA
\INSTH

\author{K.\,Abe}\INSTBJ
\author{J.\,Adam}\INSTFJ
\author{H.\,Aihara}\INSTCH\INSTHA
\author{T.\,Akiri}\INSTFH
\author{C.\,Andreopoulos}\INSTEH\INSTFC
\author{S.\,Aoki}\INSTCC
\author{A.\,Ariga}\INSTEE
\author{S.\,Assylbekov}\INSTFG
\author{D.\,Autiero}\INSTJ
\author{M.\,Barbi}\INSTE
% ...

\collaboration{The T2K Collaboration}\noaffiliation
         \begin{abstract}
        
\revcom{This paper reports a measurement by the T2K experiment of the $\nu_{\mu}$ Charged Current Quasi-Elastic (CCQE) cross section on a carbon target with the off\hyp{}axis detector}
based on the observed distribution of muon momentum ($\pMu$) and angle with respect to the incident neutrino beam ($\thetaMu$).
The flux\hyp{}integrated CCQE cross section was measured to be $\fluxAvgSigma = \fluxAvgXsecRFG$. The energy dependence of the CCQE cross section is also reported.
The axial mass, $\maccqe$, of the dipole axial form factor was extracted assuming the Smith\hyp{}Moniz CCQE model with a relativistic Fermi gas nuclear model.
\collcom{Using the absolute (shape-only) $\pMuCosThetaMu$ distribution,} 
the effective $\maccqe$ parameter was measured to be 
$\bestFitMaccqeResultShapePlusNorm$ ($\bestFitMaccqeResultShapeOnly$).
     \end{abstract}
    \maketitle
    
\section{Introduction}
\label{sec:introduction}

The Charged\hyp{}Current Quasi\hyp{}Elastic (CCQE) interaction, 
$\nu_l + n \rightarrow l^{-} + p$, 
is the dominant CC process for neutrino\hyp{}nucleon interactions at $\eNu \sim 1$~\GeV
and contributes to the total cross section in the neutrino energy range relevant for current accelerator\hyp{}based long\hyp{}baseline neutrino oscillation experiments. 
The CCQE interaction of a neutrino with a nucleon 
is a two\hyp{}body interaction and hence the initial neutrino energy can
be estimated from relatively well\hyp{}measured final state lepton kinematics
without relying on reconstruction and energy measurement of the hadronic final state. 
As the neutrino oscillation probability depends on the neutrino energy, 
the energy dependence of the CCQE cross section is critically important.
This interaction has been modeled with the Llewellyn Smith \cite{LlewellynSmith} 
(Smith\hyp{}Moniz \cite{SmithMoniz1,SmithMoniz2}) 
\revcom{formalism} for CCQE interactions on nucleons (nuclei).
Assuming a dipole axial vector form factor,  there is only one free parameter to be fixed by neutrino experiments,
the axial mass, $\maccqe$. 
This CCQE model is described in detail in \cref{sec:interactionmodel}.

Modern accelerator\hyp{}based neutrino oscillation experiments use nuclear targets 
to achieve high target masses and hence high event rates.
The use of a nuclear target introduces nuclear effects whose impact on the neutrino\hyp{}nucleon interaction must be understood.
Recent CCQE measurements 
have shown disagreement between experiments, and also between deuterium and higher atomic number nuclear targets. 
The value of $\maccqe$ extracted from neutrino\hyp{}deuterium scattering is $1.016 \pm 0.026$~\GeVoverCSq \cite{deuterium} 
and is consistent with results from pion electro\hyp{}production ($1.014 \pm 0.016$~\GeVoverCSq) \cite{pionelectroproduction}.
\collcom{The K2K experiment report $1.2\pm0.12$~{\GeVoverCSq} \cite{k2k} with an Oxygen target with peak energy $\sim $~1.2\GeV.}
The NOMAD experiment report $1.05 \pm 0.06$~\GeVoverCSq \cite{nomad} with a carbon target with $\avgEnu \sim 24$~\GeV.
The MiniBooNE experiment report $1.35 \pm 0.17$~\GeVoverCSq \cite{miniboone} with a carbon target with $\avgEnu \sim 0.8$~\GeV.
\revcom{The MINOS experiment report $1.23 \pm 0.20$~\GeVoverCSq with an Iron target with $\avgEnu \sim 2.8$~\GeV \cite{minos}.
}

These discrepancies have motivated theoretical work on more sophisticated models
including contributions from multi\hyp{}nucleon interactions which mimic the CCQE signal if the additional nucleons are not detected.
\revcom{The MINERvA experiment report excess vertex activity in CCQE data which may also indicate the presence of multi\hyp{}nucleon processes \cite{minerva}.}
Recent theoretical developments are reviewed in \cite{mec1,mec2}.
Such additional processes not only contribute to the measured total cross section, 
they can also affect the neutrino energy reconstruction. 
These new models have had success in describing
MiniBooNE CCQE data \cite{Nieves_MB,Martini_MB,Bodek_MB,Mosel_MB,RGF1} 
and T2K CC data \cite{t2knd280ccinc,Martini_T2K,SuSAv2,SuSAv3,RGF2}. 
\revcom{However, models that simultaneously describe all datasets from low and high energy experiments remain elusive
\cite{SuSAv2,SuSAv3,Nieves_highE,RGF1,RGF2}.}

\collcom{
This paper reports the first measurement by the T2K experiment of the $\nu_{\mu}$ 
CCQE cross section on a carbon target with the T2K off-axis near detector (ND280)
based on the observed distribution of muon momentum ($\pMu$) and angle with respect to the incident neutrino beam ($\thetaMu$).
Measurements of the $\pMuCosThetaMu$ distribution are analyzed within the context of 
the standard Smith\hyp{}Moniz CCQE model with a Relativistic Fermi Gas (RFG) nuclear model.
We measure the energy dependence of the neutrino cross section and extract the value of 
the $\maccqe$ parameter under this assumed model.
In addition, the flux\hyp{}integrated total CCQE cross section is also reported.  
In this analysis, no attempt is made to account for additional multi\hyp{}nucleon processes that may be present.
Thus the measured cross section should be interpreted as a measurement of a ``CCQE\hyp{}like'' cross section.
The extracted value of $\maccqe$ should be interpreted as an effective-$\maccqe$
necessary to explain the observed distribution of ``CCQE\hyp{}like'' events rather than a direct measurement of the true nucleon $\maccqe$.
The extraction of the energy dependent cross section depends 
on the $\pMu-\eNu$ dependence in the Smith\hyp{}Moniz model.
While this is an inherently model dependent approach,
such measurements have proven valuable for the development and testing of new models.
This is also the approach taken in the measurements reported above, 
so direct comparison to them is meaningful.
Given the importance of this channel, and the discrepancies between datasets, 
it is important that such measurements are repeated by several experiments with 
a variety of nuclear targets.
}

The T2K experiment and ND280 detector are described in \cref{sec:beam,sec:detector}, respectively.
The event selection is described in \cref{sec:selection}.
The neutrino interaction model used for this analysis is described in \cref{sec:interactionmodel}.
The systematic uncertainties due to neutrino flux prediction, neutrino\hyp{}nucleus interaction modeling, and detector response are described in \cref{sec:systematics}.
The measurement of the CCQE cross section
and interpretation of this result as a measurement of an effective $\maccqe$
are described in \cref{sec:xsec_measurement}.
The results are summarized and future prospects are discussed in \cref{sec:summary}.

\section{The T2K Beam}
\label{sec:beam}
The T2K experiment is an accelerator-based long-baseline neutrino oscillation experiment \cite{t2kexperiment}.
\revcom{The main goals of the experiment are the discovery and measurement of $\nue (\nuebar)$ appearance in a $\numu (\numubar)$ beam 
and precision measurements of $\numu$ disappearance.}
These are typically interpreted in the standard three-flavor model as measurements of 
the neutrino mass\hyp{}squared differences and the mixing angles of the PMNS matrix \cite{pmns1,pmns2}.
Our limited knowledge of neutrino interaction cross sections, including CCQE, forms one of the largest uncertainties
 in the measurement of oscillation parameters.
The T2K off-axis detector is an excellent laboratory for measuring CCQE interactions as backgrounds 
from inelastic scattering are suppressed due to the narrow neutrino energy spectrum.

The J-PARC main ring proton beam is extracted into the neutrino beam-line.
The proton beam energy is 30~\GeV and 
each beam spill consists of up to eight bunches 15~ns in width with 581~ns spacing between bunches.
The spill cycle has a frequency of $0.3\mhyphen0.4$Hz.
The proton beam impacts a graphite target where hadronic interactions produce mostly pions and kaons.
Charged particles are focused by three magnetic horns and enter the decay tunnel where they decay to produce the neutrino beam.
For the data presented here, the horns are operated in neutrino mode to focus $\pi^{+}$ for a high purity $\nu_\mu$ beam.
A beam dump at the end of the decay tunnel stops non-neutrino particles from reaching the near detectors.
Muon monitors beyond the beam dump are used to monitor the beam intensity and direction. An underground experiment hall containing near detectors 
designed to characterize the neutrino beam 
is located 280~m from the target.
The on-axis near detector, INGRID, consists of an array of modules that measure the beam direction and profile.   
The ND280 near detector used for this analysis lies $2.5^{\circ}$ off the beam axis.
This analysis is based on 
$2.6 \times 10^{20}$ protons on target (POT) using data collected by the ND280 detector during the first three T2K running periods
(spanning the period January 2010 - June 2012).

\begin{figure}
  \includegraphics[width=0.7\linewidth]{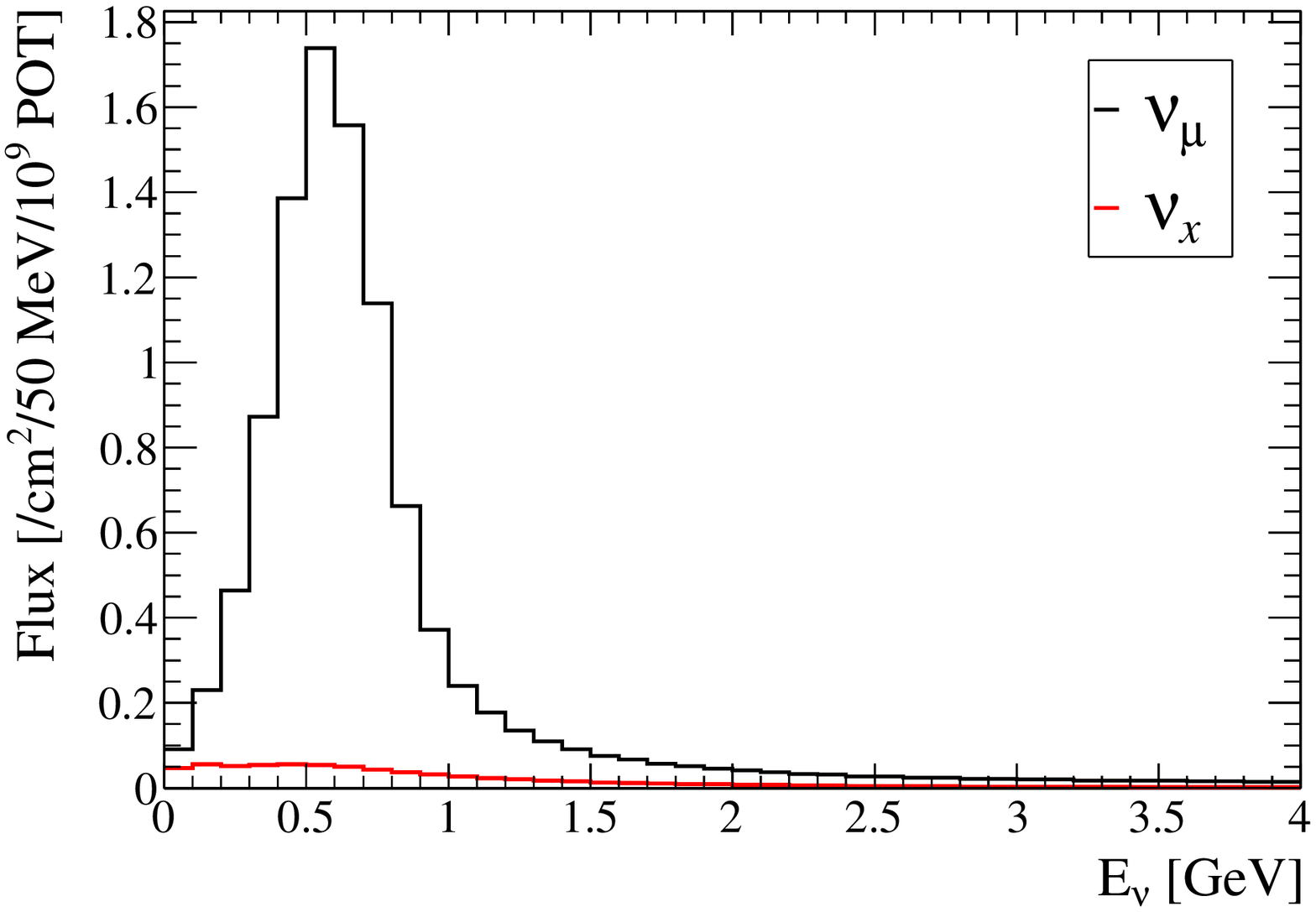}
  \caption{
  The predicted neutrino flux at the ND280. The beam is dominated by $\numu$ and is narrowly peaked at $0.6\GeV$.
  The sum of all other neutrino flavors, including anti-neutrinos, is denoted by $\nu_{x}$.
  \label{fig:nd5_run1-run2}
  }
\end{figure}

\collcom{
The predicted neutrino beam flux at the ND280 near detector peaks at $0.6\GeV$ and is shown in \cref{fig:nd5_run1-run2} \cite{beamflux}.
The neutrino beam flavor content is predicted to be $92.6\%$ $\numu$.
The $\numubar$, $\nue$, $\nuebar$ contamination are $6.2\%$, $1.1\%$, $0.1\%$, respectively.
}
The primary proton interactions with the target are simulated with FLUKA2008 
\cite{fluka1,fluka2} 
tuned to external hadron production data \collcom{such as} the NA61/SHINE experiment \cite{na61shine_1,na61shine_2}.
These simulated hadrons are propagated through the decay volume with GEANT3 \cite{geant3} with GCALOR \cite{gcalor}.

\section{The off\hyp{}axis ND280 Detector}
\label{sec:detector}

\begin{figure}
    \includegraphics[width=0.7\linewidth]{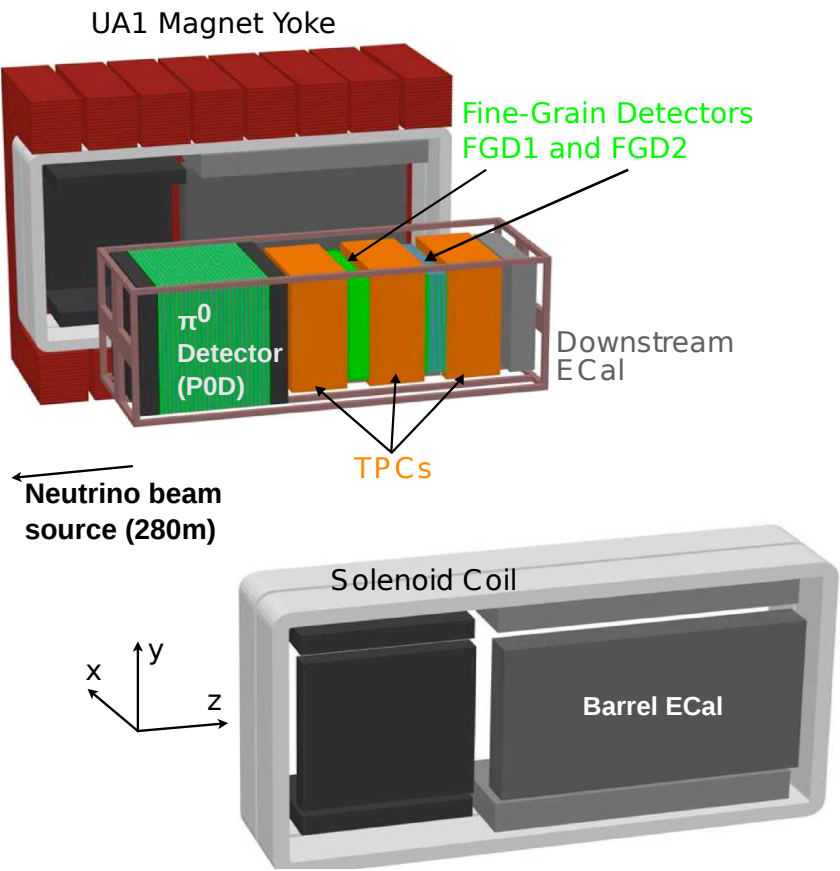}
  \caption{
\revcom{An diagram of the ND280 off axis near detector.
The diagram shows an exploded view of the ND280 detector with the surrounding magnet yoke and solenoid and e.\ m.\ calorimeter detectors
 withdrawn to reveal the tracker detector and $\pizero$ detector. The FGD and TPC are the primary sub-detectors used in this analysis 
and are described in the text.
}
  \label{fig:detector}
  }
\end{figure}

ND280 is a tracking detector located 280~m from the neutrino beam source.   
The detector sits inside the \collcom{refurbished} UA1 magnet which provides a $0.2$~T magnetic field for track sign selection and momentum measurement.
The detector is divided into two regions: a tracker \collcom{and an upstream} $\pi^{0}$ detector region (P0D) \cite{pod}.
A diagram of the ND280 detector is shown in \cref{fig:detector}.
This analysis uses events reconstructed in the ND280 tracker. 
The tracker region contains two \revcom{Fine Grained Detectors} (FGDs) and three Time Projection Chambers (TPCs).
Surrounding both the P0D and tracker regions are electromagnetic calorimeters (ECals).
The magnet is instrumented with scintillator called the Side Muon Range Detector (SMRD) 
which detects muons escaping with high angles with respect to the beam direction.

The first Fine Grained Detector (FGD1) \cite{fgd} provides target mass and track reconstruction near the interaction vertex.
It consists of layers of $9.6\mm\times9.6\mm\times1864.3\mm$ 
 plastic scintillator bars read out with wavelength shifting 
fibers into Multi\hyp{}Pixel Photon Counters (MPPCs).
\collcom{There are 30 layers with each layer containing 192 bars. 
The orientation of the layers alternates between $x$ and $y$ directions perpendicular to the neutrino beam.
This allows for three\hyp{}dimensional reconstruction of the interaction vertex.}
\collcom{The target nuclei are predominantly carbon with small fractions of other nuclei.
The composition of the target is summarized in \cref{tab:target_composition}.
}

The second FGD (FGD2) contains water layers for direct cross section measurements on water.
\collcom{Only interactions in FGD1 are used in this analysis.}
For this analysis, the fiducial volume is defined as the FGD1 volume excluding the most upstream $\xY$ layer pair
 and $5$ bars width around the edge in $\xY$ to remove external backgrounds.
 The fiducial mass is \fiducialmass.
 
The FGDs are sandwiched between three TPCs \cite{tpc}.
Each TPC consists of a box containing a gas,
consisting of Ar (95\%),
CF$_4$ (3\%) 
and iC$_4$H$_{10}$ (2\%), 
with an electric field  between the central cathode and the side anodes where drift charge is read out with micro\hyp{}mesh gas detectors (micromegas) \collcom{\cite{micromegas}}. 
\revcom{The TPC provides 3D track reconstruction, momentum measurement and sign selection from 
track curvature in the magnetic field, and particle identification (PID) from $\mathrm{d}E/\mathrm{d}x$ in the gas.
A momentum resolution of 
$\delta$(p$_\perp$)/p$_\perp \sim 0.08 $p$_\perp$ [GeV/c] 
is achieved for muons as measured from track curvature in the the magnetic field.
}
TPC tracks and FGD clusters are combined to form complete three\hyp{}dimensional tracks with the vertex position given by the most upstream FGD hit and the PID and momentum measurement given by the TPC.

\section{Event Selection}
\label{sec:selection}

CCQE interactions are selected with a cut\hyp{}based analysis which identifies events with a reconstructed $\mu^{-}$ starting within the FGD1 fiducial volume
and no reconstructed $\pi^{\pm}$.
No explicit cuts are applied to include or exclude a proton track.
In the following, CCQE is defined as the nucleon\hyp{}level interaction (defined by the model described in \cref{sec:interactionmodel}) before final state interactions 
(i.e.\ the interactions of the outgoing proton with the target nucleus).

Good data quality is required by selecting only spills where the entire ND280 detector is operational.
Tracks reconstructed in ND280 are associated with a primary proton beam bunch based on hit timing.
Tracks with a reconstructed time greater than $60\nanosec$ from the expected beam bunch mean time are rejected.
The reconstructed bunch time width is around $15\nanosec$.

Tracks starting within the FGD fiducial volume with a TPC component are selected.
The TPC component is required to contain at least \collcom{18 clusters}. 
This track quality requirement ensures that tracks are long enough to provide reliable PID and momentum information.
The $\mu^{-}$ candidate is defined as the highest momentum, negatively charged track.

Given the particle momentum, the TPC $\mathrm{d}E/\mathrm{d}x$ distribution for different particle hypotheses is known.
A muon PID variable that combines the likelihood given the observed $\mathrm{d}E/\mathrm{d}x$ for $\mu$, $e$, $\pi$ and proton hypotheses is used to select muons.
\collcom{97\% of simulated $\numu$ CCQE events that pass the previous cuts pass this muon PID cut.}

External backgrounds are removed by vetoing interactions with additional tracks starting $> 150\mm$ upstream of the $\mu^{-}$ vertex.
\revcom{This cut removes backgrounds where the true vertex is upstream (for example in the P0D or magnet region).}
Only tracks going from FGD1 forward into TPC2 are selected, as the reconstruction is currently unable to distinguish backwards-going $\mu^{-}$ from forwards\hyp{}going external backgrounds. 

This event selection selects $\numu$ CC interactions with an efficiency of 50\%
and a purity of 87\%. 
The dominant background is from neutrino interactions outside of the fiducial volume.

Two additional cuts are applied to remove events containing pions.
Events with additional tracks in the TPC are rejected as most protons from CCQE interactions stop in the FGD.
Only $14\%$ of true CCQE events selected with the CC inclusive event selection have a matching proton track reconstructed in the TPC.
Events with multiple TPC tracks are dominated by resonant pion and DIS backgrounds.
Events with delayed clusters reconstructed in the FGD are rejected to remove stopped $\pi^{\pm}$ that decay to $\mu^{\pm}$ and then $e^{\pm}$.

\begin{figure}
        \includegraphics[width=0.65\linewidth]{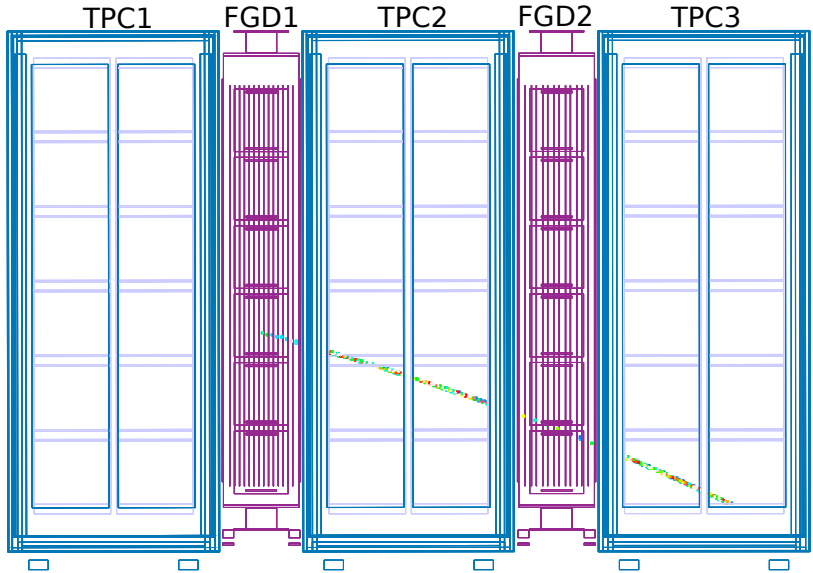}
    \caption[Generated and fit parameter distributions.]
    {
        An example CCQE candidate event.
        Only the tracker region of the ND280 is shown.
        This event shows a forward\hyp{}going muon track starting within the fiducial volume of FGD1.
        It passes through two TPCs and the second FGD before exiting the detector.         \label{fig:event_display}
    }
\end{figure}

\begin{figure}
         \subfigure[$\pMu$]{\includegraphics[width=0.75\linewidth]{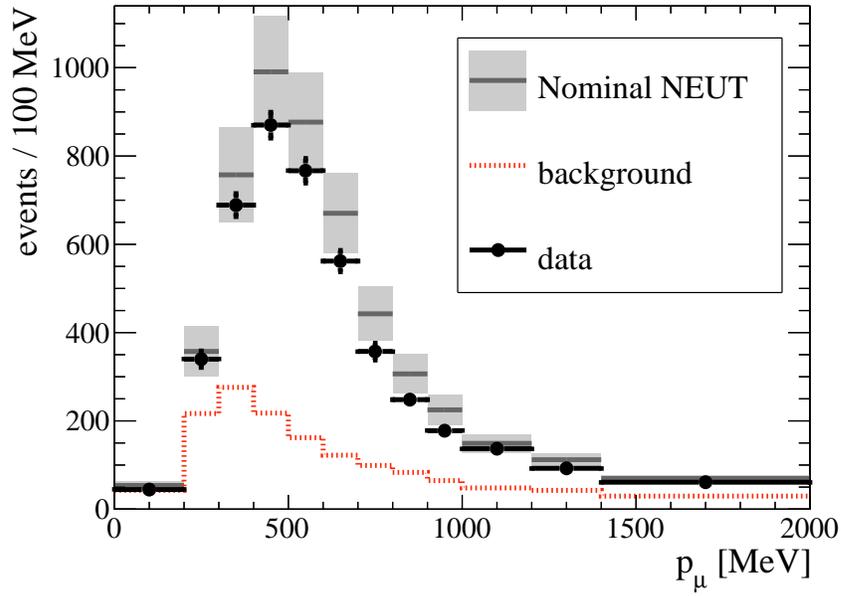}} 
     \subfigure[$\cosThetaMu$]{\includegraphics[width=0.75\linewidth]{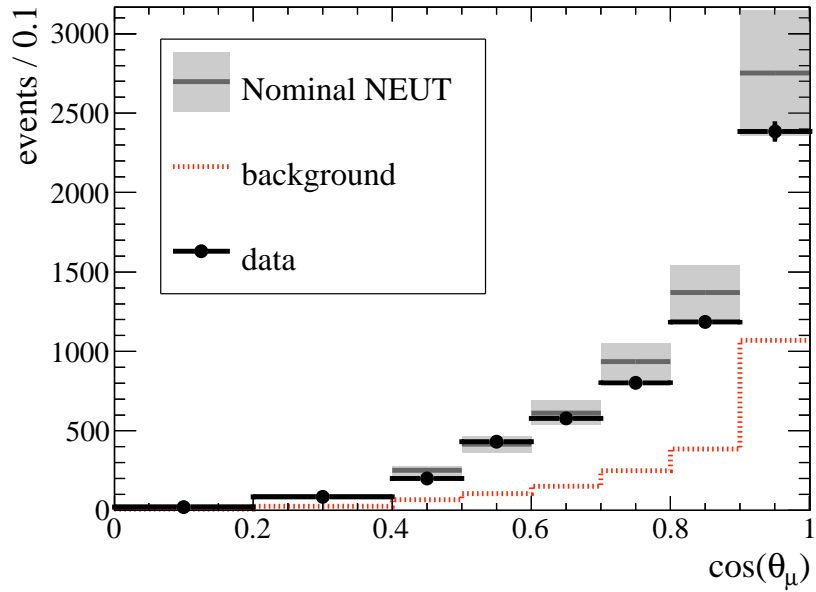}}
     \caption{
     Reconstructed muon kinematics. The background is the NEUT prediction and is dominated by CC resonant pion production.
     The gray error band on the MC prediction includes the systematic uncertainties described in \cref{sec:systematics}.
     The error bars on the black data points show the statistical uncertainties.
     } 
     \label{fig:kinematics}

\end{figure}

This event selection achieves an efficiency of $40\%$ and purity $72\%$ for true CCQE events in the model described in \cref{sec:interactionmodel}.
The dominant background is from CC resonant pion production.
The cuts and selection efficiency are summarized in \cref{tab:event_reduction_and_efficiency}.
\revcom{A total of $5841$ events were selected.}
This CCQE sample is a sub\hyp{}sample of the CC inclusive sample described in \cite{t2knd280ccinc}.
An example event display of a candidate signal event is shown in \cref{fig:event_display}.
The reconstructed muon kinematics are shown in \cref{fig:kinematics}.

\section{Neutrino Interaction Model}
\label{sec:interactionmodel}

Neutrino interactions are modeled with the NEUT Monte Carlo generator \cite{neut}.
The four\hyp{}vectors of the final state particles are propagated through a detector simulation using GEANT4.
\Cref{tab:selection} shows the fraction of simulated events passing the selection, broken down by nucleon level interaction type.
The event sample is dominated by CCQE and CC resonant single pion production. 
\collcom{The NEUT generator uses the Smith\hyp{}Moniz model for CCQE scattering with nuclear targets \cite{SmithMoniz1,SmithMoniz2}. }
In this model, the weak nucleon current is written in the most general form that conserves 
T and C symmetry and contains four form factors that must be determined by experiment.
\collcom{The electromagnetic form factors are precisely measured in electron elastic scattering experiments.
The vector form factors of the weak nucleon current are related to the electromagnetic form factors through the conserved vector current hypothesis.
Two form factors remain: the axial vector form factor, $F_A$ and the pseudo\hyp{}scalar form factor.
The pseudo\hyp{}scalar form factor can be related to the axial form factor. This leaves $F_A$ as the only independent free form factor.
}
A dipole form is assumed for $F_A$,
\begin{equation}
F_A(Q^{2}) = 
             \frac{F_A(0)}
                   {
                   \left(
                          1 + \frac{Q^{2}}{{\maccqe}^{2}} 
                   \right)^{2}
                   } .
\end{equation}
The results reported in this paper are based on a value of $\maccqe = 1.21\GeVoverCSq$ (unless otherwise stated).

In this model, nucleons are treated as non\hyp{}interacting particles bound in a potential well of binding energy, $E_B$ and momentum up to the global Fermi momentum, $p_F$.
Both of these parameters are set by electron scattering experiments.
As well as describing the nuclear initial state, 
\collcom{this also restricts the available final states through Pauli blocking}.
No additional contribution from multi\hyp{}nucleon effects are included.

\revcom{Final State Interactions (FSI) are implemented with a semi-classical cascade model.
In the cascade model hadrons are treated as classical objects under-going a sequence of independent collisions as they move through the nucleus.
This is implemented as an independent step after the initial neutrino-nucleon scattering that has the effect of altering the observed hadronic final state.
}
In the cascade model, a vertex position within the nucleus is selected from a Wood\hyp{}Saxon nucleon density distribution.
Each hadron produced in the interaction is stepped through the nucleus. 
At each step a Monte Carlo method is used to determine if an interaction occurs.
If an interaction occurs, FSI is applied to the outgoing hadrons.
This process is repeated until all hadrons escape the nucleus or are absorbed.
For pions, three interactions may occur:
inelastic scattering,
absorption
and
charge exchange.
The probabilities of these interactions depend on the pion momentum and position within the nucleus and are constrained by external pion\hyp{}nucleus scattering measurements 
\cite{pion_N_scattering_1,pion_N_scattering_2,pion_N_scattering_3,pion_N_scattering_4,pion_N_scattering_5,pion_N_scattering_6,pion_N_scattering_7,pion_N_scattering_8,pion_N_scattering_9,pion_N_scattering_10,pion_N_scattering_11,pion_N_scattering_12,pion_N_scattering_13,pion_N_scattering_14,pion_N_scattering_15,pion_N_scattering_16,pion_N_scattering_17,pion_N_scattering_18,pion_N_scattering_19,pion_N_scattering_20,pion_N_scattering_21,pion_N_scattering_22,pion_N_scattering_23,pion_N_scattering_24,pion_N_scattering_25,pion_N_scattering_26,pion_N_scattering_27}.
\revcom{Elastic scattering is neglected as it is mostly forward going with negligible momentum change.}

The Rein\hyp{}Seghal model\cite{ReinSeghal1,ReinSeghal2} is used to simulate the resonant single pion background.
In this model the production cross section of single pion plus nucleon final states is calculated by summing over 18 intermediate resonances with hadronic invariant mass $W < 2\GeVoverCSq$, including interference terms.
The nominal axial mass is $\mares = 1.16\GeVoverCSq$ and the RFG nuclear model is assumed. 
\revcom{This value is set from fits to external pion data described in \cref{sec:pionfits}.}
 
\section{Systematic Uncertainties}
\label{sec:systematics}
 
 The systematic uncertainties in this analysis can be factorized into 
 neutrino beam flux uncertainties,
 interaction model uncertainties
 and detector response uncertainties.
 A summary of the effect of these systematic uncertainties on the CCQE cross section measurement is shown in \cref{tab:systematic}. 
 
\subsection{Neutrino Beam Flux Uncertainty}
\label{sec:systematics_beam}

The neutrino beam flux uncertainty is dominated by the uncertainty on the hadron interaction model, 
including uncertainties on the total proton\hyp{}nucleus production cross section,
pion multiplicities,
and secondary nucleon production.
These uncertainties are summarized in \cref{tab:flux_errors}.
These uncertainties are derived primarily from NA61/SHINE measurements \cite{na61shine_1,na61shine_2} of pion and kaon production from proton interactions on a graphite target at the T2K beam energy.
For areas of phase space uncovered by these measurements, additional external datasets are used. These uncertainties are propagated to the T2K neutrino beam flux prediction by re\hyp{}weighting FLUKA2008 MC samples.
The total proton\hyp{}nucleus production cross section uncertainty is set to cover discrepancies between NA61 measurements \cite{na61shine_1} and other external datasets \cite{denisov} \cite{belletini} \cite{carrol}.
The uncertainty on secondary nucleon production is set by comparing the predictions of the FLUKA2008 model with external measurements \cite{eichten,allabey}.

Uncertainties in the operational conditions of the beam\hyp{}line are also considered.
These include 
the proton beam incident position on the target,
the proton beam profile \collcom{and intensity}, 
the angle between the beam and the ND280 detector,
and the horn current, field and alignment. 

For this analysis, the neutrino beam flux uncertainty is modeled as a multi\hyp{}variate Gaussian in 11 bins of true neutrino energy, including the correlation between bins.
The $\nu_{\mu}$ flux uncertainty varies between $\eNu$ bins and ranges from $10\%$ to $15\%$.
The effect on the measured flux integrated CCQE cross section is $12\%$.
Further details of the neutrino beam flux prediction can be found in \cite{beamflux}.

\subsection{Interaction Model Uncertainty}
\label{sec:systematics_xsec}

The interaction model uncertainties are motivated from comparison of the NEUT MC generator with external experimental data.
A summary of the parameters 
and their effects on the overall normalization of the selected QE\hyp{}enhanced data sample is shown in \cref{tab:priorXsecError}.
The same parameterization of the systematic error on the interaction model as 
\cite{t2knd280ccinc} is used with the exception of the parameters that affect the CCQE model.
All uncertainties that affect the CCQE normalization are removed. 

Variation of model parameters is estimated by evaluating the bin contents for model parameter variations of $0, \pm 1, \pm 2, \pm 3, \pm 4 $ and $\pm 5 \sigma$.
Cubic spline interpolation between these points is used to evaluate the change in bin contents for an arbitrary change in model parameter. 

Uncertainty on the shape of the $\pMuCosThetaMu$ distribution for CCQE events 
enters when extracting $\sigma(\eNu)$.
This is included with model uncertainties on two parameters, $\maccqe$ and $\pF$. 
For $\maccqe$, a central value of $1.21\GeVoverCSq$ is selected and the $1\sigma$ error set to $0.2$\GeVoverCSq.
This error covers the best\hyp{}fit values reported by the NOMAD and MiniBooNE experiments, as well as the $\maccqe$ value measured from deuterium scattering experiments.
The Fermi\hyp{}momentum $\pF$ parameter value and error are set from electron scattering measurements.
\collcom{For each event, with neutrino energy $\eNu$, the prediction with a parameter value $x^{\prime}$ is multiplied by the factor 
$\sigma(x^{\mathrm{nominal}}, \eNu) / \sigma(x^{\prime}, \eNu)$,
where
$\sigma(x, \eNu)$ is the total CCQE cross section calculated with the parameter value $x$.}
In both cases, the systematic uncertainty is implemented as a shape\hyp{}only parameter.
Hence, the total cross section is conserved, and only the effect of the parameter on the shape of the $\pMuCosThetaMu$ distribution is included.  

The single pion background uncertainty is set from
 analysis of 
  MiniBooNE 
 pion production 
 data 
\cite{miniboone_cc1pi0,miniboone_cc1piplus,miniboone_nc1pi0}.
Model parameter uncertainties are set on the axial vector mass for resonant interactions, $\mares$.
Normalization uncertainties are applied to the CC resonant single pion,
NC $1\pi^{0}$ and NC coherent production rates.
\label{sec:pionfits}
$W_{\mathrm{eff}}$ shape is an empirical parameter used to account for shape differences between the data and MC predictions in the NC$1\pi^{0}$ pion momentum distributions
by modifying the width of the hadronic resonance.
In the absence of direct observations of CC coherent pion production at the energy relevant for the T2K neutrino beam flux, 
a conservative estimate of 100\% is set on the CC coherent pion rate
\cite{cc_coherent1,cc_coherent2}.

``CC other'' is an $\eNuTruth$ dependent uncertainty on the inelastic
cross section excluding CC resonant single pion production.
This rate is known to $\sim 10\%$ from external datasets
\cite{ccother_ref} at 4~\GeV.
``NC other'' is a 30\% normalization uncertainty on NC elastic, NC resonant production of 
$\eta/K/\gamma$, NC DIS and NC multi\hyp{}$\pi$ production.
The contribution from NC backgrounds to this analysis is small.

The FSI uncertainty is estimated by varying parameters of the NEUT FSI model, 
changing the effective $\pi$ interaction rates within limits allowed by external 
pion\hyp{}carbon scattering data.
A covariance matrix is generated from the variation of the number of events 
in each $\pMuCosThetaMu$ bin.

\subsection{Detector Response Uncertainty}
\label{sec:systematics_det}

The detector response uncertainties are summarized in \cref{tab:priorDetError}. 
The dominant systematic uncertainties originate from the TPC momentum measurement 
and from ``out\hyp{}of\hyp{}fiducial\hyp{}volume backgrounds''.

Uncertainties in the TPC momentum scale arise due to uncertainties in the magnetic field.
The TPC momentum resolution uncertainty is estimated from analysis of through\hyp{}going muon tracks passing through multiple TPCs 
by comparing the reconstructed momentum in each TPC.

The background from out\hyp{}of\hyp{}fiducial\hyp{}volume events is $6.1\%$. These can enter the selected sample due to mis\hyp{}reconstruction.
\collcom{The dominant reconstruction effect is from muon tracks that originate from outside of the fiducial volume and pass through the FGD
but have no hits reconstructed in the bars surrounding the FGD fiducial volume.}
This is investigated by examining the hits distributions of through\hyp{}going muons that are known from the TPC measurement to pass through every layer of the FGD.
An additional $20\%$ cross section uncertainty is applied to the out\hyp{}of\hyp{}FGD events 
as they come primarily from interactions on heavier nuclei.
This is motivated by comparison of the out\hyp{}of\hyp{}FGD rates predicted by the NEUT 
and GENIE event generators as well as comparisons between data and MC of the observed rates 
in the SMRD, ECal and P0D detectors.
The total uncertainty assigned to out\hyp{}of\hyp{}fiducial\hyp{}volume interactions is up to $9\%$ depending on the reconstructed $\pMuCosThetaMu$ bin.

Uncertainties due to 
the track reconstruction efficiency, 
TPC particle ID,
Michel electron tagging,
sand interactions (i.e.\ interactions in material upstream of the ND280 detector such as the sand surrounding the pit)
and 
pile\hyp{}up
are considered.
These are studied using control samples in data including through\hyp{}going muons and cosmic muons.
The uncertainty due to secondary pion interactions
is evaluated by comparing GEANT4 
simulation with external pion\hyp{}carbon scattering datasets.
The fiducial mass uncertainty is $0.67\%$.
These systematic uncertainties are described in detail in \cite{t2knd280ccinc}.

For each source of detector response uncertainty, a covariance matrix in reconstructed $\pMuCosThetaMu$ bins is formed.
The total uncertainty was formed from the sum of the covariance matrices.

 \section{CCQE Cross Section Measurements}
\label{sec:xsec_measurement}

\subsection{Methodology}
\label{sec:fitmethod}

The CCQE cross section is extracted by applying a binned likelihood fit 
to the observed $\pMuCosThetaMu$ distribution.
The input events are grouped into bins in $\pMuCosThetaMu$ with bin edges
$\pMu=[0.0, 0.4, 0.5, 0.7, 0.9, >0.9]\GeVoverC$
and $\cosThetaMu=[-1.0, 0.84, 0.90, 0.94, 1.0]$.
For simulation, the events are generated with the NEUT MC model described in \cref{sec:interactionmodel}.
The simulated data are additionally binned in the unobserved variable $\eNuTruth$ and true interaction type.
Systematic uncertainties are accounted for by varying the bin contents with nuisance parameters.
Five parameters of interest are defined which scale the normalization of CCQE prediction in bins of true $\eNu$ with bin edges $\eNu=[0.0, 0.6, 0.7, 1.0, 1.5, >1.5]\GeVoverCSq$.
This binning is optimized to ensure a stable fit result.
The best\hyp{}fit value of these parameters of interest is used to calculate the measured CCQE cross section.
The predicted $\pMuCosThetaMu$ distribution is calculated by summing over the unobserved variables, after taking into account the bin weights.

The number of predicted events in an observed
$\pMuCosThetaMu$ bin, $i$, is,
\begin{equation}
\NumPredictedBinI =
d_{i} \displaystyle\sum_{j}^{\eNuTruth \mathrm{bins}}
f_{j}
\displaystyle\sum_{k}^{\mathrm{event\ type}}
w_{jk} x_{ijk} N^{\mathrm{MC}}_{ijk}
,
\end{equation}
where
the indices $j$ and $k$ correspond to
the unobserved quantities 
$\eNuTruth$ bins
and
true interaction type,
respectively,
$N^{\mathrm{MC}}_{ijk}$ is the number of events predicted by the nominal MC,
$d_{i}$ are weights for detector systematics,
$f_{j}$ are weights for the flux systematics, 
$x_{ijk}$ are the weights for the cross section systematics,
$w_{jk}$ are the \revcom{parameters of interest} that weight the cross section in bins of $\eNuTruth$.
These $w_{jk}$ parameters are only allowed to vary for CCQE interactions.
For all other interactions they are fixed at unity. 
The true interaction types are shown in \cref{tab:selection}. 
This categorization is used for the re\hyp{}weighting used to implement the model systematic uncertainties.

A binned maximum likelihood fit is used to find the best fit parameters.
The negative log\hyp{}likelihood ratio is minimized,
\begin{equation}
\hspace*{-10mm}
        \begin{array}{ll}
    -2 \mathrm{ln} \lambda(\bm{\Theta}) =   
    2 \displaystyle\sum^{\tiny \pMuCosThetaMu \mathrm{bins}}_{\tiny i=1}
    & \left[
          \NumPredictedBinITheta - \NumObservedBinITheta 
          \right.
           \left.
          + \NumObservedBinITheta \mathrm{ln} \frac{ \NumObservedBinITheta }{ \NumPredictedBinITheta }
          \right] \\
          & 
           + \mathrm{ln}\frac{\pi_{d}(\bm{\Theta_d})}{\pi_{d}(\bm{\Theta_d}^{\mathrm{nominal}})} 
           + \mathrm{ln}\frac{\pi_{f}(\bm{\Theta_f})}{\pi_{f}(\bm{\Theta_f}^{\mathrm{nominal}})} 
           + \mathrm{ln}\frac{\pi_{x}(\bm{\Theta_x})}{\pi_{x}(\bm{\Theta_x}^{\mathrm{nominal}})} ,    
      \end{array}
\end{equation}
where
$\bm{\Theta}$ is the set of fit parameters, and 
$\pi_{d}, \pi_{f}, \pi_{x}$ are multi\hyp{}variate Gaussian constraints on the nuisance parameters that describe the detector, neutrino beam flux and interaction model uncertainties.
The final value for the extracted $\CCQE$ cross section is calculated by multiplying the weights, $w_{jk}$,
 by the corresponding flux\hyp{}integrated cross section, $\left< \sigma_{jk} \right>_{\Phi}$, in the nominal model.
The uncertainty on the CCQE cross section is estimated with the constant $\Delta \mathrm{ln} \lambda$ method.
Studies with pseudo\hyp{}data generated from the MC model found that this method is unbiased and gives frequentist coverage.
To avoid experimenter bias, a blind analysis was performed. The analysis method was developed and validated with toy Monte Carlo studies before being applied to the data. 
 
\subsection{Energy Dependent Cross Section}
\label{sec:xsecresults}

\begin{figure}
                \includegraphics[width=0.99\linewidth]{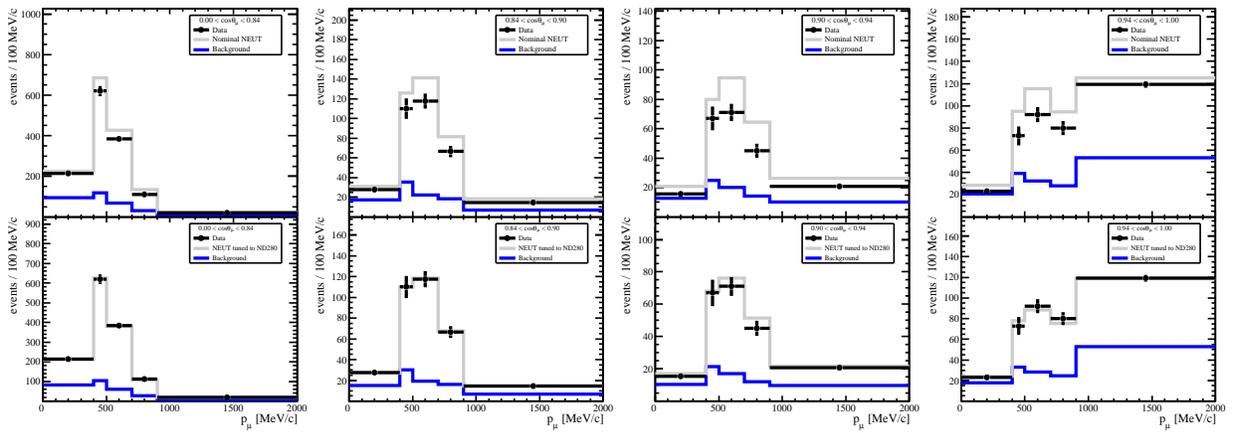} 
    \caption{
       \revcom{Data MC comparison of both the predicted (top row) and fitted (bottom row) $\pMuCosThetaMu$ distributions.        
        Each plot shows the $\pMu$ distribution for a fixed range of $\cosThetaMu$.}
        \label{fig:ptheta_before_and_after}
    }
\end{figure}

\begin{figure}
        \includegraphics[width=0.75\linewidth]{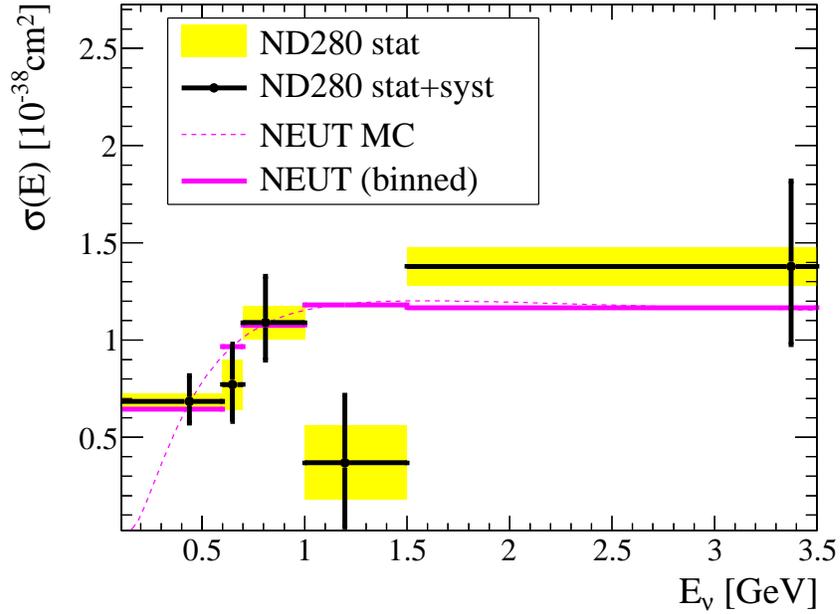}
    \caption{
    The measured CCQE energy\hyp{}dependent cross section per target neutron with statistical (band) and total (bar) errors.
    \label{fig:result_xsec}
    }
\end{figure}

\begin{figure}
        \includegraphics[width=0.75\linewidth]{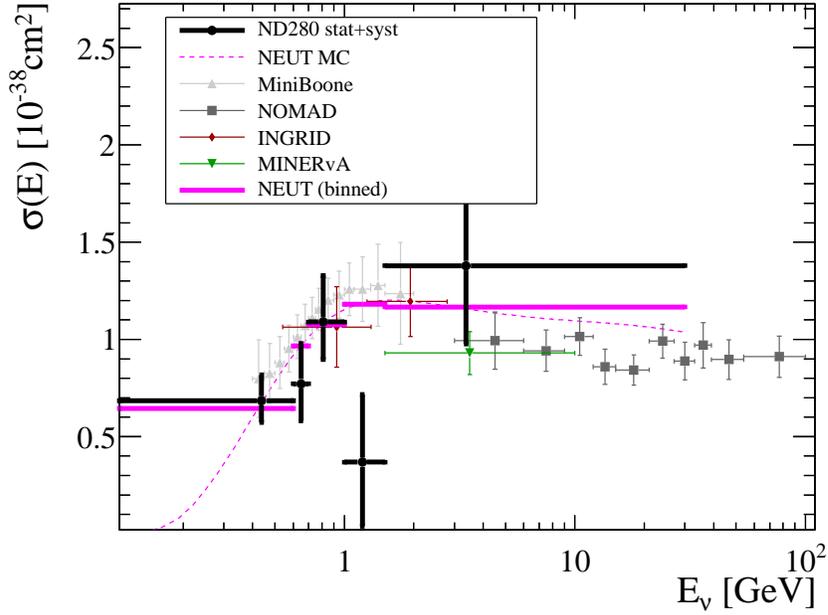}
    \caption{
    \revcom{The measured CCQE energy\hyp{}dependent cross section per target neutron compared with other experimental results \cite{miniboone,minerva,nomad,ingrid}.}
    \label{fig:result_xsec_external_data}
    }
\end{figure}

\Cref{fig:ptheta_before_and_after} shows the observed $\pMuCosThetaMu$ distribution before and after the fit.
\Cref{fig:result_xsec} shows the best\hyp{}fit CCQE cross section.
The best-fit to the ND280 data prefers a lower overall cross section than the model prediction.
In general the nominal model prediction (Smith-Moniz with RFG and $\maccqe=1.2$~\GeVoverCSq)
gives a good description of the data. 
One exception is the measurement in the range $1\textrm{-}1.5\GeV$ which differs from the model prediction at the level of $2.3\sigma$ in that bin.
However, this discrepancy is not statistically significant. 
A $\chi^{2}$ test comparing the fitted result with the nominal model prediction gives a $p$-value of 17\% 
indicating agreement between the data and the cross section model.
This is compared with measurements from other experiments in \cref{fig:result_xsec_external_data}.
There is consistency between the experiments within the current statistical and systematic uncertainties.
At low $\eNu$ the uncertainties are comparable with other experiments.
At high $\eNu$ the uncertainties increase where there is increased contamination from background processes and larger flux uncertainties.
The total fractional covariance matrix is given in \cref{tab:result_covariance}.
These data may be useful to compare with alternative models, however care should be taken to account for the model assumptions made in the extrapolation from $\pMuCosThetaMu$ to $\eNu$.

\subsection{Flux Averaged Total Cross Section}
\label{sec:fluxavgxsec}

The flux averaged CCQE cross section, $\fluxAvgSigma$, is calculated with the following equation,

\begin{equation}
\fluxAvgSigma = \frac{ 
                            \displaystyle\sum_{j}{ f_{j}\phi^{\mathrm{nominal}}_{j} w_{jk} \sigma^{\mathrm{nominal}}_{jk} }
                        }
                        {
                            \displaystyle\sum_{j} f_{j}\phi^{\mathrm{nominal}}_{j}
                        } 
                        ,
\end{equation}
where $f_{j}$ is the fitted value of the flux weight nuisance parameter,
$\phi^{\mathrm{nominal}}_{j}$ is the nominal predicted flux in $\eNuTruth$ bin $j$,
$w_{jk}$ is the fitted value of the CCQE signal cross section weight,
$\sigma^{\mathrm{nominal}}_{j}$ is the nominal CCQE cross section in bin $j$.
This is a useful quantity to calculate, since it does not rely on extrapolation from observed quantities $\pMuCosThetaMu$ to $\eNu$;
therefore
it is inherently less model-dependent.
\collcom{The fitted flux weight nuisance parameters, $f_{j}$, all remain close to their nominal values.
The choice to use the post-fit nominal values as opposed to their fitted values has a negligible effect on the final 
$\fluxAvgSigma$.
To estimate the error on the flux-averaged CCQE cross section a Monte Carlo method was used. 
A multi-variate Gaussian PDF was constructed using the best-fit $w_{jk}$ and $f_{j}$ parameter values and the post-fit covariance matrix.
A toy MC was used to generate 100000 sets of $w_{jk}$ and $f_{j}$ parameters.
These values were used to calculated a set of $\fluxAvgSigma$.
The standard deviation of this distribution was used to estimate the error on $\fluxAvgSigma$.
}
The final value from the fit to the data is {\fluxAvgXsecRFG} per target neutron.
This is in good agreement with the nominal model prediction $\fluxAvgXsecNominalNEUT$  (with $\maccqe=1.21$~\GeV).

\subsection{$\maccqe$ Extraction}
\label{sec:maqeeffective}

An alternative approach to fitting the cross section normalization 
is to directly fit the model parameters.
The axial mass parameter $\maccqe$ is varied to obtain the best fit to the observed data.
The axial mass parameter affects both the total cross section and its $Q^{2}$ dependence.
It is interesting to consider both the effect on $\maccqe$ with and without including overall normalization information in the analysis.

Two fits are performed. 
First, a version that simultaneously varies the shape and normalization 
(referred to as $\maccqe$\hyp{}norm).
Second, a version that varies only the shape 
(referred to as $\maccqe$\hyp{}shape).
The $\maccqe$ shape uncertainty described in \cref{sec:systematics_xsec} is removed from the analysis for these fits.
Therefore, the fitted values are independent of external data constraints.
The best fit values for $\maccqe$-norm ($\maccqe$\hyp{}shape) is 
$\bestFitMaccqeResultShapePlusNorm$ ($\bestFitMaccqeResultShapeOnly$).
A goodness\hyp{}of\hyp{}fit test is performed by comparing the minimized log\hyp{}likelihood ratio value 
with the expected distribution from Monte Carlo simulation.
The one\hyp{}sided $p$\hyp{}value is calculated to be $p=0.67$ $(0.68)$ for $\maccqe$\hyp{}norm ($\maccqe$\hyp{}shape)
indicating a good fit of the model to these data. 

The shape\hyp{}only fit result prefers high values of $\maccqe$ relative to the shape and normalization fit result.
As shown in \cref{sec:fluxavgxsec}, the total CCQE cross section is in good agreement, within error,
with the nominal NEUT model.
As the total CCQE cross section scales with $\maccqe$, the $\maccqeNorm$ 
parameter is constrained by the overall normalization to lie close to the nominal value.
This normalization constraint does not affect the $\maccqeShape$ fit
and hence this parameter is free to explore a wider range of the parameter space.
While higher values are preferred when using shape\hyp{}only, it should be noted that these results are consistent within their current uncertainties.
As discussed in \cref{sec:introduction}, low values of 
$\maccqe$ ($\sim 1$~\GeVoverCSq) are observed in measurements of CCQE 
interactions in neutrino deuterium scattering and pion production in $ep$ scattering. 
Note that the meaning of this effective parameter depends on the details of the QE model; 
comparison with results from other experiments should be done with care.

\section{Summary}
\label{sec:summary}

We have selected $\numu$ CCQE interactions on carbon and analyzed the observed $\pMuCosThetaMu$ distribution 
within the context of the standard CCQE interaction model with a RFG nuclear model.
The model gives a good description of the data \collcom{within the current level of statistical and systematic uncertainty.}
The flux\hyp{}integrated CCQE cross section was measured to be $\fluxAvgSigma = \fluxAvgXsecRFG$ 
in good agreement with the nominal model value of $\fluxAvgXsecNominalNEUT$ (evaluated at $\maccqe = 1.21$~\GeVoverCSq).
We have extracted the energy\hyp{}dependent CCQE cross section.
Understanding this dependence is crucial for current and future neutrino oscillation experiments.
Both results are consistent with an enhanced cross section relative to a model with $\maccqe = 1.03\GeVoverCSq$
(or equivalently this is consistent with a high effective $\maccqe$).
A direct extraction of $\maccqe$ yields 
$\bestFitMaccqeResultShapePlusNorm$ when using normalization plus shape information and 
$\bestFitMaccqeResultShapeOnly$ when using only shape information.
This observation is consistent with that observed by other experiments at similar energies
\collcom{however at the current level of uncertainty we are unable to resolve the discrepancies between NOMAD and MiniBooNE datasets.} 
\label{sec:future}
\revcom{The generators used by the T2K experiment are being actively developed.
Future analyses will include multi-nucleon effects and more advanced nuclear models.
In addition, future T2K CCQE analyses will produce a model independent measurement 
of the flux integrated differential cross section for QE-like processes as a 
function of muon kinematics.}

\begin{acknowledgments}
We thank the J-PARC staff for superb accelerator performance and the
CERN NA61 collaboration for providing valuable particle production data.
We acknowledge the support of MEXT, Japan;
NSERC, NRC and CFI, Canada;
CEA and CNRS/IN2P3, France;
DFG, Germany;
INFN, Italy;
National Science Centre (NCN), Poland;
RSF, RFBR and MES, Russia;
MINECO and ERDF funds, Spain;
SNSF and SER, Switzerland;
STFC, UK; and
DOE, USA.
We also thank CERN for the UA1/NOMAD magnet,
DESY for the HERA-B magnet mover system,
NII for SINET4,
the WestGrid and SciNet consortia in Compute Canada,
GridPP, UK.
In addition participation of individual researchers
and institutions has been further supported by funds from: ERC (FP7), EU;
JSPS, Japan;
Royal Society, UK;
DOE Early Career program, USA.
\end{acknowledgments}

    \clearpage
    
\nolinenumbers
    
\newcommand{\mcw}[1]{\multicolumn{1}{c}{#1}}

\begin{table}
        \caption{
    \collcom{
    The elemental composition of the FGD1 in the fiducial volume.
    The composition is expressed both as a fraction of the total mass of target nuclei 
    and the total number of target neutrons (for CCQE interactions).
    }
    \label{tab:target_composition}
    }
    \begin{ruledtabular}
    \begin{tabular}{ldd}
     Element & \text{Mass (\%)} & \text{Neutron (\%)} \\
     \hline
     C & $86.1$ & $92.8$ \\
     H & $7.4$ & - \\
     O & $3.7$ & $4.0$ \\
     Ti & $1.7$ & $1.9$ \\
     Si & $1.0$ & $1.1$ \\
     N & $0.1$ & $0.2$ 
    \end{tabular}
    \end{ruledtabular}
\end{table}

\begin{table}
        \caption
    {
        The predicted cumulative signal efficiency and purity at each cut level.
        \label{tab:event_reduction_and_efficiency}
    }
    \begin{ruledtabular}
    {
\begin{tabular}{ldd}
    Cut & \mcw{\text{Efficiency ($\%$)}} & \mcw{\text{Purity ($\%$)}} \\ 
    \hline
    Good quality negative track in FV & 50 & 26 \\
    TPC veto                          & 49 & 34 \\
    PID cut                           & 47 & 45 \\
    TPC-FGD multiplicity              & 40 & 67 \\
    Michel electron veto              & 40 & 72 \\
\end{tabular}
}
     \end{ruledtabular}
\end{table}

\begin{figure}
  \includegraphics[width=0.75\linewidth]{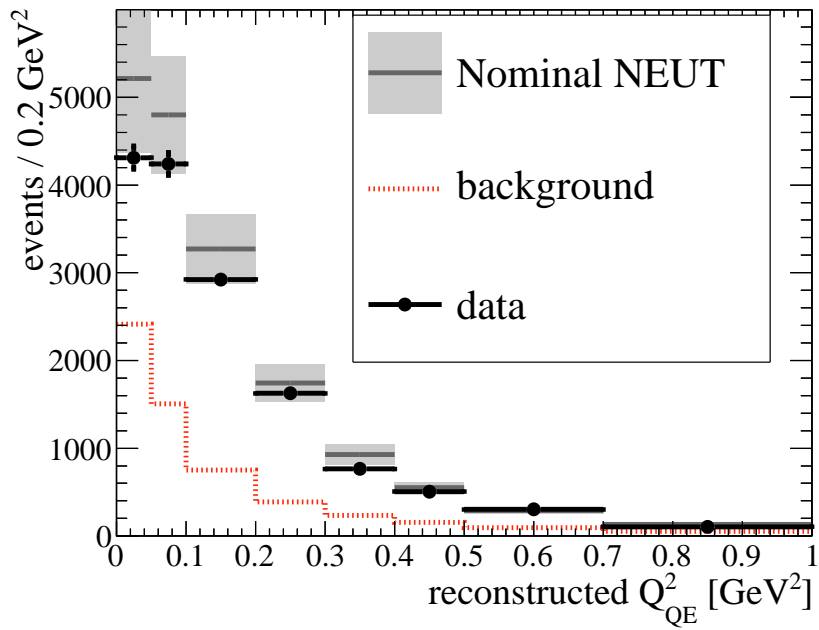}
  \caption{
    \revcom{Data MC comparison of the $Q^{2}$ distribution calculated assuming QE kinematics. 
     The gray error band on the MC prediction includes the systematic uncertainties described in \cref{sec:systematics}.
     The error bars on the black data points show the statistical uncertainties.
    }
  }
\end{figure}

\begin{table}
        \caption{
    Predicted fraction of selected events broken into interaction type.     CC other includes resonant production of multiple\hyp{}pions, 
    production of other mesons (such as $\eta$ and $K$)
    and deep\hyp{}inelastic scattering. 
    \collcom{Sand interactions occur in the material upstream of the ND280 detector such as the sand surrounding the pit and the pit walls}.
    \label{tab:selection}
    }
    \begin{ruledtabular}

\begin{tabular}{ld}
Event Type & \mcw{\text{Fraction (\%)}} \\ \hline
CCQE              & $72.0$ \\
CC$1\pi$          & $16.1$ \\
CC coherent       & $1.8$ \\
CC other          & $6.3$ \\
NC                & $2.2$ \\
Sand interactions & $1.5$ \\
\end{tabular}
     \end{ruledtabular}
\end{table}

\begin{table}
        \caption{
    A summary of the effect of each class of systematic uncertainty on the flux\hyp{}integrated CCQE cross section.
    \label{tab:systematic}
    }
    \begin{ruledtabular}
    \begin{tabular}{lc}
        Error Source & Frac. Err. on $\fluxAvgSigma$ (\%)\\
    \hline
    Detector & 4\\ 
    Neutrino beam flux & 12\\ 
    Interaction model & 4\\ 
    Statistical & 8\\ 
    \hline 
    Total & 16\\
    \end{tabular}
     \end{ruledtabular}
\end{table}

\begin{table}
        \caption{
    A summary of the neutrino beam flux uncertainties. The uncertainties shown are the uncertainty on the total $\numu$ neutrino flux.
    \label{tab:flux_errors}
    }
    \begin{ruledtabular}
    {
\begin{tabular}{ld}
   & \mcw{\text{Error (\%)}} \\ 
    \hline

Secondary nucleon production & 6.9 \\
Production cross section     & 6.4 \\
Pion multiplicity            & 5.0 \\
Kaon multiplicity            & 0.8 \\
Off-axis angle               & 1.6 \\
Proton beam                  & 1.1 \\
Horn absolute current        & 0.9 \\
Horn angular alignment       & 0.5 \\
Horn field asymmetry         & 0.3 \\
Target alignment             & 0.2 \\

    \hline
Total                        & 10.9 \\
\end{tabular}
}
     \end{ruledtabular}
\end{table}

\begin{table*}
       \caption
    {
        The pre\hyp{}fit model parameter errors and nominal values. 
        The effect of each systematic on the total normalization of the predicted number of selected events is also shown. 
        \label{tab:priorXsecError}
    }
    \begin{ruledtabular}
    {
\begin{tabular}{clccc}
Parameter & Description & Nominal value & Error & Normalisation (\%) \\ \hline
\ensuremath{M_{A}^{QE}} & axial mass for QE interactions & 1.21 GeV/\ensuremath{c^{2}} & 0.20\GeVoverCSq & 3.5\% \\
\ensuremath{p_{F}} & Fermi momentum & 217 MeV/c & $30.38$ MeV/c & 1.5\% \\ 

$x_1^{CC1\pi}$ & CC1\ensuremath{\pi} norm. ($0.0 \le E_{\nu} < 2.5$)   & 1.63  & 0.43 & 4.3\% \\
$x_2^{CC1\pi}$ & CC1\ensuremath{\pi} norm. ($E_{\nu}>2.5$) & 1  & 0.40 & 1.9\% \\\
\ensuremath{M_{A}^{RES}} & axial mass for resonant interations       & 1.16 GeV/\ensuremath{c^{2}} & 0.11 & 1.8\% \\

$W_{\mathrm{eff}}$ & modifies width of hadronic resonance      & 1 & 0.52 & 0.6\% \\
$x^{CCcoh.}$ & CC coherent norm. & 0  & 1.00 & 1.0\% \\

$x^{CCother}$ & varies other CC interactions& 0  & 1.00 & 0.6\% \\
$x^{NC1\pi}$ & NC1\ensuremath{\pi} norm.   & 1.19  & 0.43 & 0.4\% \\

$x^{NCcoh.}$ & NC coherent norm. & 1  & 0.3 & 0.4\% \\
$x^{NCother}$ & varies other NC interactions & 1  & 0.3 & 0.5\% \\

$x^{FSI}$ & FSI & - & - & 0.5\% \\

\end{tabular}
}

     \end{ruledtabular}
\end{table*}

\begin{table}
       \caption
    {
        The detector systematic errors. 
        The effect of each systematic on the total normalization of the predicted number of selected events is also shown. 
        \label{tab:priorDetError}
    }
    \begin{ruledtabular}
    {
\begin{tabular}{ccc}
Systematic Error & \mcw{\text{Normalisation Error (\%)}} & \mcw{\text{Shape Error (\%)}} \\ \hline

TPC momentum scale          & $0.1$ & $0.2 \textrm{-} 5.9$            \\ 
TPC momentum resolution     & $0.2$ & $0.0 \textrm{-} 2.0$            \\
External background         & $1.3$ & $0.4 \textrm{-} 8.9$            \\
 
Track reconstruction     & $0.6$ & $0.7 \textrm{-} 2.1$            \\
TPC PID                     & $0.02$ & $0.0 \textrm{-} 0.7$            \\

Michel tagging  & $0.6$ & $0.3 \textrm{-} 1.6$            \\

Sand interations            & $0.1$ & $0.0 \textrm{-} 1.1$            \\
Pile-up                     & $0.4$ & $0.3 \textrm{-} 1.6$            \\
Pion rescattering           & $1.4$ & $0.5 \textrm{-} 4.7$            \\

Fiducial Volume Target Mass & $0.6$ & $0.6$ \\

\end{tabular}
}

     \end{ruledtabular}
\end{table}

\begin{table}
        \caption{
    The fractional covariance matrix corresponding to the errors shown in \cref{fig:result_xsec}.
    \label{tab:result_covariance}
    }
    \begin{ruledtabular}
    {
\begin{tabular}{c|ddddc}
Energy bin (\GeV) & \mcw{$0.0\text{-}0.6$} & \mcw{$0.6\text{-}0.7$} & \mcw{$0.7\text{-}1.0$} & \mcw{$1.0\text{-}1.5$} & \mcw{$\ge 1.5$} \\ \hline 
$0.0\mhyphen0.6$ & 0.035 & 0.002 & 0.024 & 0.006 & 0.025 \\
$0.6\mhyphen0.7$ & 0.002 & 0.038 & -0.004 & 0.023 & 0.000 \\
$0.7\mhyphen1.0$ & 0.024 & -0.004 & 0.039 & -0.006 & 0.035 \\
$1.0\mhyphen1.5$ & 0.006 & 0.023 & -0.006 & 0.068 & -0.021 \\
$\ge 1.5$ & 0.025 & 0.000 & 0.035 & -0.021 & 0.102 \\
\end{tabular}
}
     \end{ruledtabular}
\end{table}

\end{document}